# Magnetic properties of Sohncke-type Pb(TiO)Cu$_4$(PO$_4$)$_4$

S. W. Lovesey [1,2,3]

[1] *ISIS Facility, STFC, Didcot, Oxfordshire OX11 0QX, United Kingdom*
[2] *Diamond Light Source, Harwell Science and Innovation Campus, Didcot, Oxfordshire OX11 0DE, United Kingdom*
[3] *Department of Physics, Oxford University, Oxford OX1 3PU, UK*

**Abstract** The enantiomorphous (chiral) crystal class of the Sohncke-type insulator Pb(TiO)Cu$_4$(PO$_4$)$_4$ permits the rotation of the plane of polarization of light (optical activity). Copper ions participate in antiferromagnetic order below a temperature ≈ 7 K, with magnetoelectric and piezomagnetic effects permitted. Lattice and magnetic symmetries of Pb(TiO)Cu$_4$(PO$_4$)$_4$ are fully incorporated in calculated resonant x-ray Bragg diffraction patterns that are successfully compared with existing limited measurements on paramagnetic and magnetically ordered Pb(TiO)Cu$_4$(PO$_4$)$_4$ [Misawa *et al.* Phys. Rev. B **103**, 174409 (2021)]. Specifically, there is additional intensity in the ordered phase to a Bragg spot (a chiral signature) from circular polarization in the primary beam of x-rays. It is created by Cu axial magnetic dipoles, with the prospect of future experiments revealing interference between magnetic dipoles and charge-like (time-even, Templeton-Templeton) quadrupoles. Polar and magnetic (parity- and time-odd) Dirac quadrupoles and octupoles are potentially strong sources of diffraction when the reflection vector is parallel to the unique direction in the tetragonal lattice.

## I. INTRODUCTION

The lattice symmetry of crystals permits optical activity (rotation of the plane of polarization of light) in only 15 of the 32 crystal classes. There are 11 enantiomorphous classes and the 4 non-enantiomorphous classes [1-3]. The Sohncke-type lattice of the compound Pb(TiO)Cu$_4$(PO$_4$)$_4$ of immediate interest belongs to the enantiomorphic crystal class 422 [4]. Copper ions form a non-centrosymmetric antiferromagnetic structure using the crystal class 4'22' below a Néel temperature ≈ 7 K [4]. Bulk magnetic properties include magnetoelectric (ME) and piezomagnetic (PM) effects. In more detail, a Landau free energy compatible with 4'22' includes non-linear contributions in electric (E) and magnetic (H) fields. The latter are forbidden in crystal classes that contains anti-inversion ($\overline{1}'$), e.g., centrosymmetric compounds Cr$_2$O$_3$ (trigonal, magnetic crystal class $\overline{3}'$m' [5]), GdB$_4$ (tetragonal, 4/m'm'm' [6]) and Co$_2$V$_2$O$_7$ (monoclinic, 2/m' [7]). Anti-inversion in the magnetic crystal class imposes PT-symmetry, and it protects x-ray diffraction patterns from circular polarization (helicity) in the primary beam. The protection is absent in the crystal class 4'22'. We calculate the intensity circular polarization brings to a Bragg spot in the diffraction pattern, and refer to it as a chiral signature (ϒ) for Pb(TiO)Cu$_4$(PO$_4$)$_4$.

In our view, the standing of the magnetic properties of Pb(TiO)Cu$_4$(PO$_4$)$_4$ is in disarray after publication of a faulty analysis of resonant x-ray Bragg diffraction patterns [8]. We revisit diffraction amplitudes for Pb(TiO)Cu$_4$(PO$_4$)$_4$ to argue our position. To this end, we use the established magnetic structure [4], and a theory of resonant x-ray Bragg diffraction derived with standard Racah algebra for atomic multipoles [9-11]. The theory is compatible with tried and tested sum-rules in dichroic signals [12, 13]. This desirable attribute is not fully realized in a phenomenological theory used by Misawa *et al*. [8] that contains free parameters and a constraint to cylindrical Cu site symmetry [14-17]. In consequence, diffraction amplitudes are not compatible with the full magnetic symmetry of Sohncke-type $Pb(TiO)Cu_4(PO_4)_4$. Moreover, the particular application of the phenomenological theory contains a non-trivial error [8]. According to our calculations, available diffraction data are not manifestations of crystal chirality (handedness) and magnetic quadrupole interference [8]. Returning to our well-established theory of resonant x-ray diffraction, electronic multipoles therein can be estimated using an atomic wavefunction for the resonant ion [18-20], and simulations of electronic structure [21, 22].

## II. LATTICE AND MAGNETIC SYMMETRIES

The parent lattice for $Pb(TiO)Cu_4(PO_4)_4$ is $P42_12$ (tetragonal, No. 90, crystal class 422) [4]. Copper ions $Cu^{2+}$ occupy general sites 8g devoid of symmetry, and coordinates x ≈ 0.267, y ≈ 0.981, z ≈ 0.401 [4]. A neutral screw axis $2_1$ in the Sohncke lattice is achiral while the atomic structure around the axis is chiral [23]. (Of the 65 Sohncke lattices primitive ones are chiral and centred ones are not. Orthorhombic and lower symmetry lattices do not contain one of 11 enantiomorphous pairs and the related space groups are achiral.) Below a temperature ≈ 7 K, axial copper magnetic dipoles possess antiferromagnetic order described by the magnetic space group $P4'2_12'$ (No. 90.97, crystal class 4'22' [24]). This space group is equivalent to the irreducible representation $\Gamma_2$ deduced from an analysis of a magnetic neutron diffraction pattern [4]. The lattice and magnetic structures of $Pb(TiO)Cu_4(PO_4)_4$ are not centrosymmetric, not polar and not compatible with ferromagnetism. Magnetic order has a propagation vector = (0, 0, 0). Notably, the magnetic structure of an altermagnet possesses a zero propagation vector, but it is a collinear centrosymmetric antiferromagnet [25]. For such an altermagnet, a chiral signature and a PM effect are allowed, and a linear ME effect is forbidden [26].

## III. RESONANT X-RAY BRAGG DIFFRACTION

X-ray diffraction patterns gathered on crystalline materials can contain Bragg spots that do not exist in patterns created by spheres of atomic charge located at points on the particular lattice. Their inherent weakness is off-set by tuning the energy of primary x-rays from a synchrotron source to a specific atomic resonance [14-20]. The weak Bragg spots are not indexed by Miller indices for the lattice symmetry, i.e., they are space-group forbidden. Departures from spheres of atomic charge are usually labelled by components of an axial charge-like quadrupole that are invariant with respect to operations in the symmetry of sites occupied by the resonant ions (Neumann's Principle [27, 28]). Specifically, acentric sites such as those occupied by Cu ions in $Pb(TiO)Cu_4(PO_4)_4$ can harbour polar (parity-odd) multipoles that are charge-like (time-even) or magnetic (time-odd Dirac multipoles) [10, 20, 29, 30].

States of x-ray polarization, Bragg angle θ, and the plane of scattering are shown in Fig. 1. The x-ray scattering length in the unrotated channel of polarization σ → σ', say, is modelled by $(\sigma'\sigma)/D(E)$. In this instance, the resonant denominator is replaced by a sharp oscillator $D(E) = \{[E - \Delta + i\Gamma/2]/\Delta\}$ with the x-ray energy E in the near vicinity of an atomic resonance $\Delta$ of total width $\Gamma$, namely, $E \approx \Delta$ and $\Gamma \ll \Delta$. The cited energy-integrated scattering amplitude $(\sigma'\sigma)$, one of four amplitudes, is studied using standard tools and methods from atomic physics (Racah algebra) and crystallography [10, 31]. In general, a vast spectrum of virtual intermediate states makes the x-ray scattering length extremely complicated [19, 20]. It can be truncated following closely steps in celebrated studies by Judd and Ofelt of optical absorption intensities of rare-earth ions [32, 33]. An intermediate level of truncation used here reproduces sum rules for axial dichroic signals created by electric dipole - electric dipole (E1-E1) or electric quadrupole - electric quadrupole (E2-E2) absorption events [9, 12, 13]. A similar analysis exists for polar absorption events such as E1-E2, and E1-M1 where M1 is the magnetic moment [34, 35].

Electronic degrees of freedom of Cu ions are encapsulated in spherical multipoles $\langle O^K_Q \rangle$, with rank K and (2K + 1) projections in the interval $-K \leq Q \leq K$. Cartesian and spherical components $Q = 0, \pm 1$ of a vector $\mathbf{n} = (a, b, c)$, for example, are related by $a = (n_{-1} - n_{+1})/\sqrt{2}$, $b = i(n_{-1} + n_{+1})/\sqrt{2}$, $c = n_0$. A complex conjugate of a multipole is defined as $\langle O^K_Q \rangle^* = (-1)^Q \langle O^K_{-Q} \rangle$, meaning the diagonal multipole $\langle O^K_0 \rangle$ is purely real. The phase convention for real and imaginary parts labelled by single and double primes is $\langle O^K_Q \rangle = [\langle O^K_Q \rangle' + i \langle O^K_Q \rangle'']$. Whereupon, $\langle O^1_a \rangle = -\sqrt{2} \langle O^1_{+1} \rangle'$ and $\langle O^1_b \rangle = -\sqrt{2} \langle O^1_{+1} \rangle''$. For the most part, the present study appeals to an E1-E1 absorption event at the Cu $L_3$ absorption edge ($E \approx 930$ eV) [8]. The reduced matrix element of parity-even multipoles $\langle T^K_Q \rangle$ with K = 0, 1, 2 and a time signature $\sigma_\theta = (-1)^K$ appears in Ref. [10].

## IV. DIFFRACTION AMPLITUDES AND CHIRAL SIGNATURE

An electronic structure factor,

$$\Psi^K_Q = [\exp(i\boldsymbol{\kappa} \cdot \mathbf{d}) \langle O^K_Q \rangle_\mathbf{d}], \tag{1}$$

specifies a Bragg diffraction pattern for a reflection vector $\boldsymbol{\kappa}$ defined by integer Miller indices $(h, k, l)$. The implied sum in Eq. (1) is over 8g sites $\mathbf{d}$ used by Cu ions.

For a reflection vector $(0, 0, l)$ and generic multipoles $\langle O^K_Q \rangle$ [24],

$$\Psi^K_Q(0, 0, l) = [1 + (-1)^Q + 2 \sigma_\theta \cos(\pi Q/2)] [\gamma \langle O^K_Q \rangle + \gamma^* (-1)^K \langle O^K_{-Q} \rangle], \tag{2}$$

with $\gamma = \exp(i2\pi l z)$. The first factor in Eq. (2) imposes $\Psi^K_Q(0, 0, l) = 0$ for odd Q. The structure factor is also zero for Q = 0 and a time signature $\sigma_\theta = -1$. In consequence, axial magnetic dipoles and anapoles (Dirac dipoles) do not participate in diffraction for a wavevector $(0, 0, l)$. Evidently, there are no forbidden reflections of the type $(0, 0, l)$. The parity of $\langle O^K_Q \rangle$ is absent in Eq. (2), so the electronic structure factor is correct for axial and polar absorption absorption events. One finds $(\pi'\sigma) = 0$ for an axial E1-E1 event [11]. The amplitude for rotated polarization using Dirac multipoles $(\pi'\sigma)_{12}$ can be different from zero, and multipoles are denoted by $\langle G^K_Q \rangle$ for an E1-E2 event [10, 11, 20]. Setting $\sigma_\theta = -1$ and $K = 2, 3$ in Eq. (2),

$$(\pi'\sigma)_{12} = \sqrt{(2/15)} \cos^2(\theta) [\gamma' \cos(2\psi) \{\langle G^2_{+2}\rangle' + 2\sqrt{2}\langle G^3_{+2}\rangle''\}$$

$$+ i\gamma'' \sin(2\psi) \{-\langle G^2_{+2}\rangle'' + 2\sqrt{2}\langle G^3_{+2}\rangle'\}]. \qquad (3)$$

The crystal b axis is in the plane of scattering for $\psi = 0$. Intensity of a Bragg spot $|(\pi'\sigma)_{12}|^2$ is a four-fold periodic function of the azimuthal angle $\psi$, in keeping with $(0, 0, l)$ parallel to the unique direction in the tetragonal lattice. The intensity in question is absent in the paramagnetic phase of Pb(TiO)Cu$_4$(PO$_4$)$_4$, and an E1-E1 event does not contribute to the $(0, 0, l)$ diffraction amplitude in the rotated channel of polarization.

The conditions on Q in Eq. (2) are not universal for Sohncke-type lattices. For orthorhombic P2$_1$2$_1$2$_1$ (No. 19) and cubic P2$_1$3 (No. 198), for example, the condition on Q is even $(l + Q)$ [36]. Space group No. 19 describes many molecular compounds. NaClO$_3$ and NaBrO$_3$ use the cubic No. 198 lattice, and possess the same chirality yet opposite senses of optical rotation.

The electronic structure factor for $(h, 0, 0)$,

$$\Psi^K_Q(h, 0, 0) = \langle O^K_Q \rangle [\{\alpha + \alpha^* (-1)^Q\} + \sigma_\theta (-1)^h \exp(-i\pi Q/2) \{\beta + \beta^* (-1)^Q\}]$$

$$+ \langle O^K_{-Q} \rangle (-1)^{h+K} [\{\alpha + \alpha^* (-1)^Q\} + \sigma_\theta (-1)^h \exp(i\pi Q/2) \{\beta + \beta^* (-1)^Q\}], \qquad (4)$$

possesses space group forbidden reflections, i.e., $\Psi^K_0(h, 0, 0) = 0$ for even K and odd $h$. Spatial phase factors $\alpha = \exp(i2\pi hx)$ and $\beta = \exp(i2\pi hy)$.

In practice, our chiral signature $\Upsilon$ is the measured difference in intensities of a Bragg spot observed with oppositely handed primary x-rays. Thus, $\Upsilon$ and XMCD signals are alike with regard to polarization requirements. For $(h, 0, 0)$ with odd $h$ and an E1-E1 absorption event,

$$\Upsilon(h, 0, 0) = [(\pi'\pi)^*(\pi'\sigma)]'' = (1/\sqrt{2}) \cos(\theta) \sin(2\theta) [- A_0 A_1$$

$$+ 2 \sin(2\psi) \{A_0 (\alpha' + \beta') \langle T^2_{+2}\rangle'' - A_1 (\alpha'' \langle T^2_{+1}\rangle' - \beta'' \langle T^2_{+1}\rangle'')\}], \qquad (5)$$

where the definition of $\Upsilon$ anticipates $(\sigma'\sigma) = 0$. Axial magnetic dipoles in Eq. (5) are $A_0 = [4 (\alpha' + \beta') \langle T^1_c\rangle]$ and $A_1 = [2\sqrt{2} (\alpha'' \langle T^1_b\rangle - \beta'' \langle T^1_a\rangle)]$. The crystal c axis is normal to the plane of scattering at the start of the azimuthal angle $\psi = 0$. Diffraction amplitudes for a reflection vector $(h, 0, 0)$ with odd $h$ are;

$$(\sigma'\sigma) = 0, \qquad (\pi'\pi) = \sin(2\theta) [(i/\sqrt{2}) \cos(\psi) A_0 + \sin(\psi) A_1], \qquad (6)$$

$$(\pi'\sigma) = \cos(\theta) \, [- (i/\sqrt{2}) \sin(\psi) \, A_0 + \cos(\psi) \, A_1 - i\cos(\psi) \, B_1 - \sin(\psi) \, B_2],$$

with purely real quadrupoles $B_1 = [4\{\alpha''\langle T^2_{+1}\rangle' - \beta''\langle T^2_{+1}\rangle''\}]$ and $B_2 = [4(\alpha' + \beta') \langle T^2_{+2}\rangle'']$. Rotated amplitudes $(\pi'\sigma)$ and $(\sigma'\pi)$ are related by a change in sign of the magnetic dipoles.

Misawa et al. [8] report Bragg spot intensities and simulation data [22] for a reflection vector $(h = 1, 0, 0)$ and an azimuthal angle $\psi = 0$. Notably, they observe a chiral signature in the magnetic phase of Pb(TiO)Cu$_4$(PO$_4$)$_4$ with a sample temperature = 6 K (Néel temperature ≈ 7 K). According to Eq. (5) the chiral signature $\Upsilon(1, 0, 0)$ at $\psi = 0$ is a product of magnetic dipoles. Misawa et al. [8] arrive at a different interpretation of the chiral signature depicted in Fig. 3b of their paper. However, their equivalent result Eq. (17) for a chiral signature is incompatible with the magnetic space group P4'2$_1$2'. An error in Eq. (17) leads to an erroneous factor $\cos(2\psi)$ multiplying the product of dipoles. Moving to the paramagnetic phase and dipoles $A_0 = A_1 = 0$, experimental data and overlaid simulation results [22] displayed in Fig 2d [8] are consistent with an intensity $|(\pi'\sigma)|^2$ from Eq. (6) if quadrupoles therein satisfy $|B_1| \gg |B_2|$. Misawa et al. [8] arrive at a similar conclusion.

## V. CONCLUSIONS

In summary, we have investigated the magnetic properties of Sohncke-type Pb(TiO)Cu$_4$(PO$_4$)$_4$ using scattering amplitudes for resonant x-ray Bragg diffraction by the 3d-transition metal ion [10, 23]. Our results comply with the established magnetic space group P4'2$_1$2' (No. 90.97, crystal class 4'22' [24]) [4]. Regarding the theory of resonant x-ray diffraction, the spectrum of virtual intermediate states in the photon scattering length is truncated using the method pioneered by Judd and Ofelt for optical absorption intensities of rare-earth ions [31, 32, 33, 37]. Sum-rules for parity-even dichroic signals are embedded in the electronic multipoles of the resonant ions [10, 12, 13].

A predicted magnetic chiral signature agrees with limited diffraction patterns using circular polarization (helicity) in the primary beam of x-rays [8]. Future experiments can test changes to the signature with rotation of the crystal about the reflection vector (an azimuthal angle scan). Measured and calculated paramagnetic diffraction agree when one set of quadrupoles dominate. According to a calculation of the amplitude in the rotated channel of photon polarization (denoted by $(\pi'\sigma)$ in Fig. 1), there is no diffraction enhanced by an electric dipole - electric dipole (E1-E1) event for a reflection vector parallel to the unique direction of the tetragonal lattice. Diffraction by Dirac multipoles is allowed for this special reflection vector, however. We predict that anapoles (Dirac dipoles) are forbidden. Intensity enhanced by the polar electric dipole - electric quadrupole (E1-E2) absorption event exposes Dirac quadrupoles and octupoles.

Misawa et al. [8] interpret their diffraction patterns for Pb(TiO)Cu$_4$(PO$_4$)$_4$ with a phenomenological theory that does not comply with the magnetic space group [4, 14-17]. Not surprisingly, their conclusions and our conclusions do not agree for the most part.

**ACKNOWLEDGEMENTS** Dr K. S. Knight and Dr D. D. Khalyavin provided ongoing support with lattice and magnetic symmetries. Communications with Dr Y. Tanaka clarified work in Ref. [8].

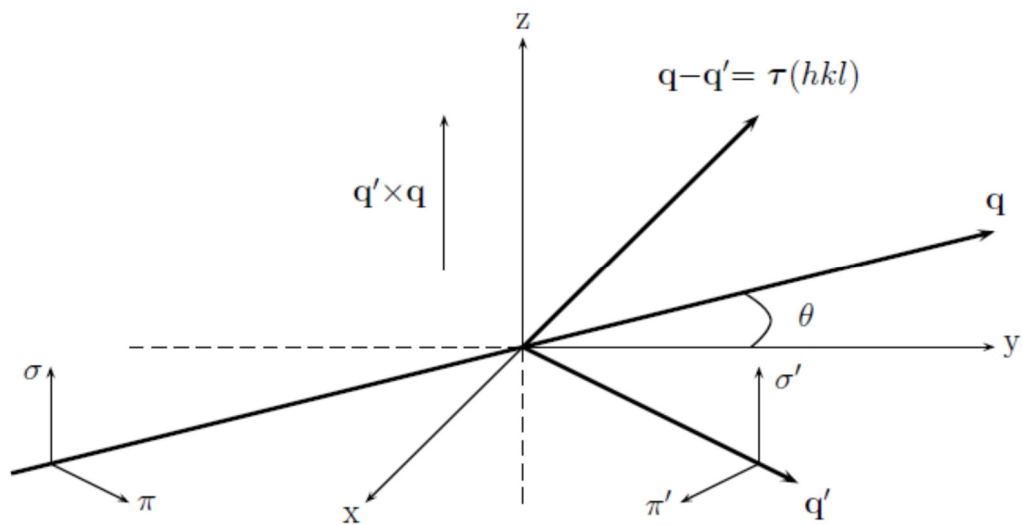

**FIG. 1.** Primary (σ, π) and secondary (σ', π') states of x-ray polarization. Corresponding wavevectors **q** and **q'** subtend an angle 2θ. The Bragg condition for diffraction is met when **q** − **q'** coincides with a reflection vector ($h$, $k$, $l$). Lattice vectors (a, b, c) and the depicted Cartesian (x, y, z) coincide in the nominal setting of the crystal, and the beginning ψ = 0 of an azimuthal angle scan (rotation of the crystal by an angle ψ about the reflection vector).